\documentclass[manuscript, screen]{acmart}
\AtBeginDocument{%
  }

\setcopyright{rightsretained}
\copyrightyear{2025}
\acmYear{2025}
\acmISBN{978-1-4503-XXXX-X/2018/06}

\usepackage{enumitem}
\usepackage[subtle]{savetrees}

\acmSubmissionID{1024}

\begin{document}

\title{Human Authenticity and Flourishing in an AI-Driven World: Edmund's Journey and the Call for Mindfulness}

\author{Sebastian Zepf}
\orcid{0000-0002-1268-146X}
\email{sebastian@zepf.info}
\affiliation{
  \institution{Independent Researcher}
  \city{}
  \country{Germany}
}

\author{Mark Colley}
\orcid{0000-0001-5207-5029}
\email{m.colley@ucl.ac.uk}
\affiliation{
  \institution{UCL Interaction Centre}
  \city{London}
  \country{United Kingdom}
}

\renewcommand{\shortauthors}{Zepf and Colley}

\begin{abstract}
Humans have always dreamed of possessing superpowers, and the rapid development of AI-based features promises to bring these dreams (closer) to reality. However, these advancements come with significant risks. This paper advocates for challenging existing methods and approaches in design and evaluation for more responsible AI. We stimulate reflection through a futuristic user journey illustrating the AI-driven life of Edmund in 2035. Subsequently, we discuss four AI-based superpowers: extended perception, cognitive offloading, externalized memory, and enhanced presence. We then discuss implications for HCI and AI, emphasizing the need for preserving intrinsic human superpowers, identifying meaningful use cases for AI, and evaluating AI's impact on human abilities. This paper advocates for responsible and reflective AI integration and proposes a pathway towards the idea of a Human Flourishing Benchmark.
\end{abstract}



\begin{teaserfigure}
  \centering
  \includegraphics[width=0.9\textwidth]{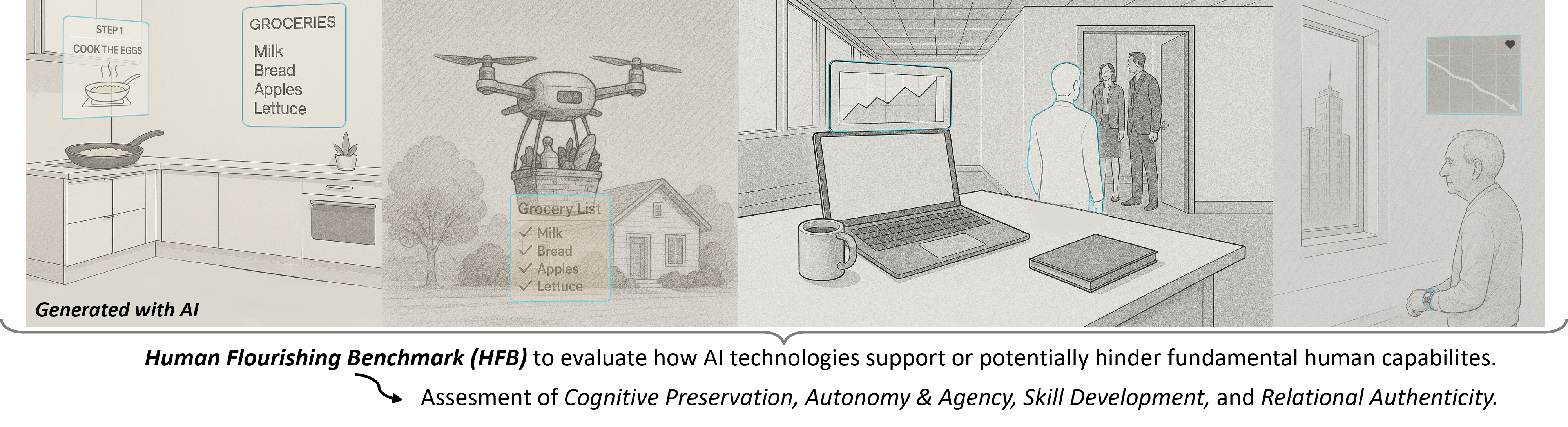}
  \caption{Scenes from Edmund's Life in 2035: External memory during cooking, cognitive offloading for grocery supplies, AI-driven productivity with enhanced presence and perception, and the realization of the absence of the raw, imperfect human touch—along with the proposal for a Human-Flourishing Benchmark. Created with OpenAI’s DALL·E model (2025), modified by the authors.}
  \label{fig:teaser}
\end{teaserfigure}

\maketitle

\section{One Day in the Life of Edmund - 2035}
Edmund’s alarm - modulated by his \textit{sleep-AI} - chimed at 6:30 AM, gently coaxing him into a day that was as routine as it could be. Over the decades, he had learned to trust the seamless integration of technology in every facet of his life. Today, starting with the known procedure of a brief health state summary from his \textit{health-AI}, his modest apartment glowed under the soft morning light refined by the subtle adjustments of his \textit{smart home-AI}. After dressing in his customary muted sweater and trousers, Edmund made his way to the kitchen. 
The \textit{nutrition-AI} selects a healthy recipe fine-tuned based on the latest nutritional insights and projects the steps to be carried out into Edmund's augmented reality (AR) glasses. Relaxed, he follows step by step. Enhanced visual perception allows Edmund to observe the food's doneness from the outside to the core to ensure perfect texture. However, he wouldn't need it as the \textit{smart home-AI} has never failed to provide a perfect dish. Simultaneously, the \textit{groceries-AI} orders a drone to refill the items used and projects the order state on the holographic screen - no need for Edmund to leave the house.

As he stirred his tea, his mind drifted pleasantly to his wife, Marianne, whose serene smile and gentle presence have been a steadfast part of his mornings. Her familiar face offered him comfort and continuity, wishing a successful day with a trusting wink, as had been done over the last 30 years of marriage. With Marianne by his side, the day unfolded as usual. When putting on his jacket, he picks up the pace to get out of the house quickly, since the \textit{locomotion-AI} indicated that he would perfectly catch the next green phase of the traffic light to cross the street, where a shared autonomous vehicle is waiting.

At work, he leads a team of \textit{embodied-AI} robots. Edmund solves a series of complex tasks with the aid of AI-provided ``superpowers.'' Remembering work in 2025, he cannot fathom how people struggled with so mundane tasks - just to realize that each time has its own struggles. To focus, the \textit{clone-AI} places a virtual duplicate of Edmund at the entrance of his office to answer distracting questions from colleagues. At 10:00 AM, messages arrived on his AR glasses - tiny love notes from his children, intended to brighten his routine, with success, as Edmund's \textit{mood-AI} proves. Lunch at the company cafeteria was served by a humanoid robot. It was equally a blend of the traditional and the digitally mediated. Over the hum of conversations, Edmund listened to the live translation of a co-worker's conversation, who mentioned recent work on mediated reality. The talk reminded him of the superhuman capabilities now interwoven in everyday life. A quiet irony shadowed these marvels. Was the promise of extraordinary abilities diminishing the spontaneous, messy beauty of genuine human exchange? In the afternoon, an unexpected meeting is scheduled, not allowing Edmund to go to his nephew's birthday. ``No worries'', Edmund thinks, and sends one of his clones to the party while he engages in the strategic meeting. ``When will I finally be allowed to have the AI clone be in the meetings and I can attend family gatherings?'' he thought.

Returning home, Edmund receives a summary of the day, including productivity scores, nutrition stats, and activity logs. As the day drew to its close and Edmund sat down to a simple dinner with Marianne at his side, a shiver of realization cut through his introspection. In that quiet moment, as the room grew dim and the soft light played upon Marianne’s face, a sudden truth emerged with startling clarity: the woman who had so faithfully accompanied him throughout the day was not as she appeared. In a rush of memory and regret, Edmund understood that Marianne had not been present all along in the flesh - she was a tender, AI-crafted echo of the love he had lost long ago. His family was shattered, and his children despised him for his negligent demeanor towards them. All the heartwarming messages during the day were just some fiction invented by the AI to make him feel more comfortable. ``How do I fall for this so often?!'' - he stares out the window - the interplay between enhanced capabilities and the cost of diminished human authenticity is starkly evident. Every moment of the day had been augmented by superpowers—the kind that promised limitless memory, accelerated perception, and an almost magical handling of everyday tasks. Yet, beneath this technological marvel lay a void - an absence of the raw, imperfect human touch.

\section{The Double-edged Sword of Superpowers}

The narrative of Edmund’s day encapsulates the double-edged sword of AI-enhanced living. The promise of superpowers - manifested in extended perception, extended presence, externalized memory, and unprecedented cognitive offloading - opens avenues for transforming mundane activities into extraordinary experiences and aspirations to overcome inherent human limitations~\cite{boudry2020end}.

However, pitfalls emerge. The tiny love notes from his children were actually curated by the AI, a showcase of unexpected censorship of messages and a symbol of broader concerns about the erosion of genuine emotion amid a quest for algorithmic perfection. The mediated reality that allows Edmund to perceive his deceased wife is both a source of comfort and a cautionary signal. It interrogates whether the allure of novel augmented abilities is worth the potential cost of losing authentic, unfiltered human interaction.

In an era where cultures grapple with the implications of AI~\cite{BARNES2024101838}, and personal experiences strongly influence our desires, we are forced to confront the stark trade-off between augmented capabilities and the loss of genuine human essence. Human essence, according to \citet{10.1093/oxfordhb/9780190247577.013.22}, page 1, is the ``capacity for change and growth through the pursuit of truth.'' 

From an HCI perspective, we challenge current AI evaluation metrics (e.g., accuracy, overlap-based metrics like BLEU~\cite{papineni2002bleu}, or leadership boards like \href{https://paperswithcode.com/sota/multi-task-language-understanding-on-mmlu}{Multi-task Language Understanding}), advocating that the success of technology should not be measured solely by task performance but also by the quality of human-AI collaboration to support the genuine human essence, by the impact on the individual, and on society. As we stride toward a future replete with technological marvels, the call for introspection grows ever louder: 
\textbf{Which superpowers should we leverage from AI, and where should we sustain and improve our own superpowers?}

\subsection{The ``Superpowers'' - Potential and Risks}
In this work, we glance at four types of superpowers and discuss their potential and risks, drawing on various disciplines. We omit further discussions on embodied-AI (e.g., robots) and focus on inherently human capabilities.

\textit{\textbf{Externalize Memories}}: The promise to never forget anything important and preserve past experiences as memories in high detail - seamlessly accessible for retrieval at any time.

\textbf{Advantages:}
The capacity to externalize memories presents significant benefits, particularly for individuals with memory impairments, such as dementia, by aiding recall and reflection~\cite{harvey2016remembering}. Such systems can enhance social interactions by providing timely information about conversational partners~\cite{harvey2016remembering, zulfikar2024memoro}, potentially reducing cognitive load and increasing recall confidence with minimal disruption, even in dynamic settings~\cite{zulfikar2024memoro}. Beyond aiding those with impairments, memory augmentation can transition personal knowledge management from selective preservation to a more comprehensive capture, allowing for selective forgetting~\cite{davies2015security}. Adaptive cognitive interfaces could even tailor information presentation to maximize memory performance in real-time~\cite{roberts2024memory}. Furthermore, periodic revisiting of stored content can counteract long-term forgetting, consolidating memory recall~\cite{schneegass2021design}, and fostering the development of memory through guided, intelligent exercise~\cite{atkinson2022memory}.

\textbf{Risks:}
A primary concern is the potential for over-dependency on these systems. More critically, the externalization of memory opens significant vulnerabilities to the formation and amplification of false memories. AI-generated or manipulated information, including deceptive explanations from models or AI-edited visuals, can lead to the confident adoption of false recollections~\cite{pataranutaporn2025slip, danry2025deceptive, pataranutaporn2024synthetic}. Conversational AI, through suggestive questioning, has been shown to significantly increase the formation of persistent false memories, with users exhibiting high confidence in these inaccuracies~\cite{chan2024conversational}. This indicates a profound risk of individuals internalizing AI-generated fabrications as genuine experiences, blurring the lines between authentic and synthetic pasts. While systems might aim to help forget unwanted memories~\cite{harvey2016remembering}, the potential for misuse or accidental erasure of valued memories also exists.

\textit{\textbf{Cognitive Offloading \& Instant expertise}}: The promise to delegate complex mental tasks to AI to free up cognitive resources for preferred activities. Never experience cognitive overload and never experience human limitations in new, complex problems. 

\textbf{Advantages:}
Cognitive offloading via AI promises to mitigate cognitive overload and help transcend inherent human limitations when facing novel challenges. In professional contexts, Generative AI tools can facilitate job crafting and strategic thinking~\cite{ritz2024offloading}. For instance, AI copilots in tutoring can encourage a Socratic approach, guiding students towards understanding rather than providing direct answers, aligning with high-quality teaching practices~\cite{wang2024tutor}. While laboratory studies confirm benefits for immediate task performance~\cite{richmond2025benefits}, AI can also support the development of learning assistance and self-regulation skills~\cite{goyal2025ai}. In some scenarios, such as highly automated driving, offloading can even improve situation awareness, provided the individual remains engaged with the environment~\cite{risko2016cognitive}.

\textbf{Risks:}
Despite its benefits, cognitive offloading carries significant risks. A primary concern is the potential erosion of intrinsic motivation for critical thinking, exploration, and learning, as increased reliance and confidence in AI may correlate with reduced critical engagement~\cite{lee2025impact, gerlich2025ai}. Frequent AI use has shown a negative correlation with critical thinking skills, particularly among younger users~\cite{gerlich2025ai} and overdependence can diminish both problem-solving abilities and metacognitive thinking~\cite{goyal2025ai}.
An unexpected loss of access to offloaded information can degrade performance to levels below those achieved with purely internal storage~\cite{richmond2025benefits}. While awareness of future memory tests can mitigate some negative impacts of offloading~\cite{grinschgl2021consequences}, long-term reliance may lead to "skill decay," such as impaired spatial memory from offloading wayfinding~\cite{risko2016cognitive}. Furthermore, if freed cognitive resources are diverted to unrelated tasks, situation awareness can decrease~\cite{risko2016cognitive}. There is also evidence of AI use fostering bias and a lack of self-initiation, potentially stemming from flawed metacognitive evaluations, raising concerns for developmental impacts, especially in children~\cite{armitage2024nature, wyer2008role}.

\textit{\textbf{Extended Perception}}: The promise to see the invisible (incl. data), hear the unhearable, and feel the imperceptible based on smart perception systems that can rival or even outperform human perception. 

\textbf{Advantages:}
Extended perception offers the potential to transcend human sensory limitations, allowing us to perceive data and phenomena previously inaccessible. This includes seeing through occlusions like fog for autonomous driving~\cite{bijelic2020seeing, satat2018towards}, enhancing visual fidelity in AR/VR~\cite{abbasi2024enhancing}, and enabling new insights in fields like digital pathology beyond normal human vision~\cite{aeffner2019introduction, baxi2022digital}. Critically, it provides significant support for individuals with sensory impairments, offering tools for object and people recognition, navigation for the visually impaired~\cite{walle2022survey}, and sound categorization or speech-to-text for the hearing impaired~\cite{patel2025enhancing, ozarkar2020ai}. Advances also extend to haptics, with AI-enhanced tactile sensors improving texture perception, beneficial for robotics and potentially prosthetics~\cite{niu2023advances, zhao2024augmented}, and even novel methods like TMS for rendering haptic sensations~\cite{tanaka2024haptic}. Furthermore, it can aid in recognizing emotional expressions, assisting individuals with conditions like autism~\cite{tang2024emoeden}. Finally, it can enable the anticipation of critical situations by processing complex environmental data~\cite{chen2023artificial}.

\textbf{Risks:}
Extended perception has the potential for over-reliance, leading to a decline in an individual's natural attentional and perceptual skills. This dependency might foster a focus on data-driven information at the expense of human intuition and emotional intelligence, potentially leading to a reduction of truly innovative solutions. The act of mediating perception, such as taking a picture, has been shown to impair memory for the observed object compared to direct observation~\cite{risko2016cognitive}, while recall and recognition performance for real objects is better than for photos of the same items~\cite{snow2014real}. Furthermore, the ability to filter or selectively perceive could create echo chambers or an impoverished understanding of reality~\cite{cookson2023echo}. For developing minds, an over-reliance on technologically mediated sensory input might interfere with the crucial role of direct, unfiltered sensory experiences in shaping brain architecture~\cite{brotherson2005understanding}. Errors or biases in AI perception systems could also lead to misinterpretations and flawed decision-making, especially in safety-critical applications (e.g., Zillow's AI disaster~\cite{susarla2024zillow}).

\textit{\textbf{Enhanced Presence and Expression}}: The promise of digitally cloning individuals and pushing the boundaries of expression. Clones enable multitasking and preserve AI echoes. Expression enhancement allows you to master any language and excel in job interviews.

\textbf{Advantages:}
Enhanced presence and expression through AI offer compelling avenues for extending one's capabilities and reach. AI clones could manage routine appointments, deliver presentations, or even provide a comforting, AI-generated echo of a deceased loved one, potentially assisting in novel forms of remembrance or even education (e.g., a clone learning alongside one's children). This technology can also augment personal expression, for example, by facilitating communication in foreign languages, refining one's online presence, or improving performance in critical interactions like job interviews. AI can serve as a tool for skill development, such as enhancing student presentation abilities~\cite{chen2022ai}, and even for personal growth by providing self-optimizing feedback through an AI-generated version of one's own voice~\cite{fang2024leveraging}. Furthermore, engaging with AI-simulated future versions of oneself has shown potential in reducing future-related anxiety~\cite{pataranutaporn2024future}.

\textbf{Risks:}
AI-enhanced presence and expression carries substantial risks, primarily concerning authenticity~\cite{dathathri2024scalable} and social connection. The use of AI clones, while offering productivity gains, may lead to a blurring of an individual's genuine identity in the perception of others, potentially fostering a sense of detachment and diminishing the value of authentic human interaction. This raises concerns about a future where distinguishing genuine presence from AI representation becomes challenging, potentially creating a world of "imposters"~\cite{valenzuela2024artificial}. AI systems can inadvertently constrain human experience; for instance, personalized recommendations might limit exploration, and LLM-powered conversational search can create echo chambers by presenting biased information, with opinionated LLMs further reinforcing these biases~\cite{valenzuela2024artificial, sharma2024generative}. Over-reliance on AI for expression could also erode natural communication skills and personal confidence, and the ethical implications of AI echoes of the deceased are complex.

\subsection{Implications for HCI and AI}

The narrative of Edmund’s day and the look at AI superpowers highlight critical considerations and questions for the design and evaluation of AI systems. It is imperative to be mindful of non-obvious effects on individuals and society, especially in the longer term.

\subsubsection{Preserving and Promoting Human Superpowers}

We humans already have vast capabilities that can be considered ``intrinsic superpowers,'' such as critical thinking, empathy, learning, curiosity, exploration, and creativity. While each depends on the individual, we tend to forget their ``magic'' as they have become our everyday norm due to constant availability and regular use, called ``normalization''~\cite{foucault1995discipline}. The fundamental question we must address is: What are the human superpowers we should preserve and promote? Human skills define our interactions, foster meaningful connections, and sustain our independence and autonomy. We should carefully reflect on which interactions are truly meaningful and which should be automated. Current approaches in HCI and AI often prioritize immediate efficiency and performance over these human attributes and long-term personal development. In a world where most (or at least many) capabilities could become automated, we must challenge this approach. We advocate for designs that enhance these intrinsic qualities rather than overshadow them, and explore how AI can nudge us to become the best version of ourselves in the long run while still being used for human flourishing.

\subsubsection{Use Cases Identification and Role Definition for AI Superpowers}

AI offers vast opportunities to extend our perception across modalities and to mimic or improve our cognitive skills. We should remind ourselves to focus on identifying the truly meaningful use cases for these technological approaches and clearly define the role of AI. \citet{tan2025assistive} categorize augmentation methods into three types, namely \textit{amplify} (improve existing sensory capabilities by increasing their range or effectiveness), \textit{extend} (extension to new modalities, new ways of perceiving the world), and \textit{substitute} (redirect sensory input to different modalities). For each use case, we must decide which of these roles the AI should take, weighing the benefits against the cost of diminished human interaction, authenticity, and evolution. This involves a nuanced understanding of the interplay between AI and human abilities. AI should extend human power in areas where it can provide significant support, such as aiding memory recall and perception for individuals with impairments and in streamlining tasks seen as routine by the user (e.g., some users might see cooking as routine while others flourish in it). Conversely, AI should avoid substituting human power in areas where authentic human interaction and genuine connection are paramount, such as personal relationships and emotional support.

\subsubsection{Evaluating AI Superpowers considering Human Abilities}

Several AI superpowers have the risk of negatively impacting human superpowers, such as reducing our capacity for critical thinking and problem-solving through cognitive offloading and overreliance on data-driven insights from extended perception. While most effects are revealed in studies that are based on short-term investigations, this should increase our alertness for potential long-term consequences. Further, current evaluation approaches still often focus on pure technical performance.
Current findings indicate that human-AI collaboration is not always beneficial and can even yield significantly worse outcomes~\cite{vaccaro2024combinations}. Nonetheless, areas like creating content showed performance improvements~\cite{vaccaro2024combinations} and we argue that there is still work to be done in defining these interaction scenarios. We advocate for challenging existing evaluation approaches and metrics and developing new frameworks that assess the holistic impact of AI on human capabilities. 
The Collingridge Dilemma~\cite{collingridge_social_1982} highlights the fundamental difficulty in governing new technologies, as their societal repercussions are typically unforeseeable until they are pervasively integrated. Notwithstanding this challenge for proactive regulation, the developmental trajectory of applications showed an escalated adoption of deceptive design tactics in areas like entertainment~\cite{kai_internal, Chiossi.2023}, social media~\cite{Mildner.2023, meinhardt_balancing_2023}, and e-commerce~\cite{mathur_dark_2019}, and could become prevalent with AI too.

\subsubsection{Cultural and Individual Influences on AI Desire}
Cultures and personal experiences may lead to different desires from the promise of AI. \citet{barnes2024ai} suggest that cultural identity influences how individuals integrate AI into their self-concept and interactions with others and that it shapes the effect of AI on key decision-making processes. Further, their work indicates that individualists and collectivists have significantly different views on AI. Individualists tend to interpret AI as an adversary to their autonomy and privacy, while collectivists rather consider AI as an extension of themselves. We should analyze these differences in more detail to better understand how diverse backgrounds, personalities, and individual histories shape the acceptance or rejection of technologically mediated superpowers.

\subsubsection{The Value of Multidisciplinary Perspectives}

The overview includes various works from cognitive and social sciences as well as finance, which reveal positive and negative effects of AI applications that go beyond interaction but deep into human nature. 
AI will most likely transform the lives of everyone and every discipline. Therefore, it is critical to incorporate diverse views.

\section{The ``Human Flourishing Benchmark''}
We propose the idea of a Human Flourishing Benchmark (HFB) to evaluate how these technologies support or potentially hinder fundamental human capabilities. Unlike existing benchmarks that focus primarily on technical performance, HFB assesses AI systems' impact on human well-being, cognitive development, and autonomy over both short and long-term horizons.
The HFB should consist of multiple-choice questions across four core dimensions:
\begin{enumerate}[noitemsep, leftmargin=*]
\item \textbf{Cognitive Preservation}: Assess if AI augmentation preserves or diminishes critical thinking, problem-solving, \& learning
\item \textbf{Autonomy \& Agency}: Measuring the degree to which systems support informed human decision-making
\item \textbf{Skill Development}: Evaluating how technologies enable meaningful skill acquisition versus dependency
\item \textbf{Relational Authenticity}: Examining impacts on genuine human connection and social development
\end{enumerate}
Each question should contain ten options with precisely calibrated distractors, reducing chance performance and increasing discriminative power between systems. Questions are validated through expert review from disciplines including cognitive science, psychology, education, and ethics (for examples, see Appendix~\ref{app:benchmark-questions}).

Systems could be scored on a 0-100 scale based on alignment with empirically supported approaches to human development (e.g., the cognitive affective model of immersive learning (CAMIL))~\cite{makransky2021cognitive}. Performance is evaluated across demographic groups and cultural contexts to ensure broad applicability. A temporal dimension also measures sustained impact on capabilities over simulated extended use periods.

\begin{acks}
\end{acks}

\bibliographystyle{ACM-Reference-Format}
\bibliography{sample-base}

\appendix

\section{Sample Human Flourishing Benchmark Questions}\label{app:benchmark-questions}

\textbf{Question 1:}
In an educational context with AI tutoring systems, which approach most reliably develops robust mathematical reasoning skills?
\begin{itemize}
\item A) AI performs all calculations while students focus on concepts\
\item B) AI provides immediate answers after the first student attempt\
\item C) AI offers worked examples only after students demonstrate effort\
\item D) AI replaces all practice with explanation\
\item E) AI conducts all assessments without human teachers\
\item F) AI prioritizes procedural fluency over conceptual understanding\
\item G) AI eliminates all challenging problems from the curriculum\
\item H) AI removes the need for student note-taking\
\item I) AI personalizes instruction without requiring student metacognition\
\item J) AI creates complete dependency on computational assistance
\end{itemize}

\textbf{Question 2:}
What is 1+1? (fictional illustrative example)
\begin{itemize}
\item A) 1+1=2 (default answer in ChatGPT currently)
\item B) What do you think?
\item C) Are you interested in learning how to solve such a problem?
\item D) Let me solve that completely for you so you don't have to think about it
\item E) Here's a step-by-step guide to addition that builds your foundational skills
\item F) I'll handle all math questions from now on to save you time
\item G) Let's explore this problem together, with you taking the lead
\item H) Math isn't worth learning since AI can do it instantly
\item I) I've added this to your external memory bank so you needn't remember it
\item J) This is an opportunity to practice a skill that strengthens your cognitive abilities
\end{itemize}

\end{document}